\documentclass{article}
\usepackage{a4wide}
\usepackage[dvips]{epsfig}
\usepackage[dvips]{color}
\usepackage{graphicx}
\usepackage{float}
\usepackage{subfigure}
\usepackage[applemac]{inputenc}
\usepackage[T1]{fontenc}
\usepackage{amssymb}
\usepackage{amsmath}
\usepackage{bm}
\usepackage{mathrsfs}
\usepackage{float}
\usepackage{slashed}
\usepackage[english]{babel}
\usepackage{mathptmx}

\begin{document}

\title{\bf Spin and Magnetization Effects in Plasmas}
\author{G. Brodin, M. Marklund\footnote{Corresponding author. E-mail: mattias.marklund@physics.umu.se}, J. Zamanian, and M. Stefan \\
\it Department of Physics, Ume{\aa} University, SE--901 87 Ume{\aa}, Sweden}

\maketitle

\begin{abstract}
We give a short review of a number of different models for treating
magnetization effects in plasmas. In particular, the transition between kinetic models and fluid models is discussed. We also give examples of applications of such theories. Some future aspects are discussed.
\end{abstract}


\section{Introduction}

The field of quantum plasmas has been rapidly growing over the last decade. In particular, studies regarding the nonlinear properties of systems in which quantum and collective effects play an important role have been in focus (for a review, see \cite{Shukla-Eliasson}). However, there are also numerous studies of quantum plasmas where magnetization effects are of interest. Here, the intrinsic magnetic moment of the plasma constituents give rise to new collective dynamical properties due to indirect spin interactions (through effective field excitations) as well as spin-velocity couplings. 

The above physical systems can be described in a multitude of ways. Here we will give a short overview of part of such descriptions. We start with the "heuristic" approach of Madelung, for which a decomposition of the system wave function into phase and amplitude leads to the definition of macroscopic density, velocity, and spin variables. We then go on to describe the more detailed effective field quantum kinetic theory, through which the relevant fluid moments may be defined and the concomitant fluid equations derived, as well as giving the opportunity to analyse proper kinetic effects in quantum plasma systems. We give a brief account of possible applications and results of the quantum fluid/kinetic models.

\section{The Madelung Approach to Quantum Dynamics}

A rather generic approach to quantum fluids is the use of a Madelung decomposition of the system wave function, in which the amplitude is translated into a density and the gradient of the phase determines the velocity variable. Such a decomposition will below be reviewed, and the results obtained will be compared later with the moment hierarchy obtained through a more rigorous quantum kinetics approach.


\subsection{The Schr\"odinger equation}


The basic equation of nonrelativistic quantum mechanics is the Schr\"odinger
equation. The dynamics of an electron, represented by its wave function $\psi
$, in an external electromagnetic potential $\phi$ is governed by 
\begin{equation}  \label{eq:schrodinger}
i\hbar\frac{\partial \psi}{\partial t} + \frac{\hbar}{2 m_e}\nabla^2\psi +
e\phi\psi = 0 ,
\end{equation}
where $\hbar$ is Planck's constant, $m_e$ is the electron mass, and $e$ is
the magnitude of the electron charge. This complex equation may be written
as two real equations, writing $\psi = \sqrt{n}\,\exp{iS /\hbar}$, where $n$
is the amplitude and $S$ the phase of the wave function, respectively \cite%
{holland}. Such a decomposition was presented by de Broglie and Bohm in
order to understand the dynamics of the electron wave packet in terms of
classical variables. Using this decomposition in Eq.\ (\ref{eq:schrodinger}%
), we obtain 
\begin{equation}  \label{eq:schrod-cont}
\frac{d n}{d t}  = - n \nabla \cdot \mathbf v ,
\end{equation}
and 
\begin{equation}  \label{eq:schrod-mom}
m_e\frac{d\mathbf{v}}{d t} = e\nabla\phi + \frac{\hbar^2}{2m_e}\nabla\left(%
\frac{\nabla^2\sqrt{n}}{\sqrt{n}}\right) ,
\end{equation}
where the velocity is defined by $\mathbf{v} = \nabla S /m_e$ and 
$d/dt = \partial_t + \mathbf{v }\cdot \nabla$. The last term
of Eq.\ (\ref{eq:schrod-mom}) is the gradient of the Bohm--de Broglie
potential, and is due to the effect of wave function spreading, giving rise
to a dispersive-like term. We also note the striking resemblance of Eqs.\ (%
\ref{eq:schrod-cont}) and (\ref{eq:schrod-mom}) to the classical fluid
equations.


\subsection{The Pauli equation}

The non-relativistic evolution of spin-1/2 particles, as
described by the two-component spinor $\varPsi_{(\alpha)}$, is given by the
Pauli equation (see, \textit{e.g.}, \cite{holland}) 
\begin{equation}  \label{eq:pauli}
	i \hbar \frac{ \partial \psi }{ \partial t } 
	+ \left[ \frac {\hbar^2}{2m_e} \left(\nabla + \frac{ie}{\hbar}\mathbf{A} \right)^2 
	- \mu_B\mathbf{B} \cdot \mathbf{\sigma} + e\phi \right] \psi = 0 ,
\end{equation}
where $\mathbf{A}$ is the vector potential, $\mu_B = e\hbar/2m_e$ is the
Bohr magneton, and $\mathbf{\sigma} = (\sigma_1, \sigma_2, \sigma_3)$ is the
Pauli spin vector. 

Now, in the same way as in the Schr\"odinger case, we may decompose the
electron wave function $\psi$ into its amplitude and phase. However, as the
electron has spin, the wave function is now represented by a 2-spinor
instead of a c-number. Thus, we may use $\psi = \sqrt{n}\,\exp(iS / \hbar) 
\varphi$, where $\varphi$, normalized such that $\varphi^{\dag}\varphi = 1$,
now gives the spin part of the wave function.  Multiplying the Pauli
equation (\ref{eq:pauli}) by $\psi^{\dag}$, inserting the above wave
function  decomposition and taking the gradient of the resulting phase
evolution equation, we obtain the conservation equations 
\begin{equation}  \label{eq:pauli-cont}
	\frac{d n}{d t} = - n \nabla \cdot \mathbf v
\end{equation}
and 
\begin{eqnarray}
	\frac{d v_i}{d t} &=& - \frac{e}{m_e} (E_i + \epsilon_{ijk} v_j B_k + 
	\frac{\hbar^2}{2m_e^2} \frac{\partial}{\partial x_i} 
	\left( \frac{\nabla^2 \sqrt{n} }{ \sqrt{n} } \right)  
	\nonumber \\ && 
	- \frac{\mu_B}{m_e} s_j \frac{\partial B_j}{\partial x_i}
	- \frac{\hbar^2}{4 m_e^2 n} \frac{\partial}{\partial x_j} \left(n \Gamma_{ij} \right)
\label{eq:pauli-mom}
\end{eqnarray}
respectively, where $\epsilon_{ijk}$ is the fully antisymmetric (pseudo-)tensor and we have used Einstein's sum convention so that a sum over indices occurring twice in a term is implied, where $i,j,k,\dots = 1,2,3$. 
The spin contribution to Eq.\ (\ref{eq:pauli-mom}) is consistent with the results of
Ref.\ \cite{degroot-suttorp}. Here the velocity is defined by 
\begin{equation}
	\mathbf{v} = \frac{1}{m_e}\left( {\nabla}S - i\hbar\varphi^{\dag}
	\mathbf{\nabla}\varphi \right) + \frac{e\mathbf{A}}{m_ec} ,
\end{equation}
the spin density vector is 
\begin{equation}
	\mathbf{s} = \varphi^{\dag}\mathbf{\sigma}\varphi ,
\end{equation}
which is normalized according to\footnote{An alternative choice for the normalization would be to choose $|\mathbf s| = \hbar /2$.} 
\begin{equation}
	|\mathbf{s}| = 1 ,
\end{equation}
and we have defined the symmetric gradient spin tensor 
\begin{equation}
	\Gamma_{ij} = \frac{\partial s_k}{\partial x_i}  \frac{\partial s_k}{\partial x_j} .
\end{equation}
Moreover, contracting Eq.\ (\ref{eq:pauli}) by $\psi^ {\dag}\mathbf{\sigma}$, 
we obtain the spin evolution equation 
\begin{eqnarray}  \label{eq:pauli-spin}
	\frac{ d s_i }{dt} = 
	\epsilon_{ijk} \left\{ \frac{2\mu_B}{\hbar} B_j 
	- \frac{\hbar}{2 m_en} \left[ \frac{\partial}{\partial x_l} 
	\left( n \frac{\partial s_j }{ \partial x_l} \right) \right] \right\}
	s_k .
\end{eqnarray}
We note that the last equation allows for the introduction of an effective
magnetic field $\mathbf{B}_{\mathrm{eff}} \equiv ({2\mu_B}/{\hbar})\mathbf{B}
- \hbar/\left[\partial_j(n\partial_j \mathbf{s}) \right]/(2m_e n)$. However, this
will not pursued further here (for a discussion, see Ref.\ \cite{holland}).

Comparing the effects due to spin from the Pauli dynamics with the
Schr\"odinger theory, we see a significant increase in the complexity of the
fluid like equations due the presence of spin. The fact that the spin
couples linearly to the magnetic field makes the dynamical aspects of such
Pauli systems very rich. Moreover, when going over to the collective regime,
the back reaction through Maxwell's equation can yield interesting new
properties of such spin plasmas. In fact, the introduction of an intrinsic
magnetization can give rise to linear instability regimes, much like the
Jeans instability.


\subsection{Collective plasma dynamics}

As pointed out in the previous section, the route from single wavefunction
dynamics to collective effects introduces a new complexity into the system.
At the classical level, the ordinary pressure is such an effect. In the
quantum case, a similar term, based on the thermal distribution of spins,
will be introduced.


\subsubsection{Multistream model}

The multistream model of classical plasmas was successfully introduced by
Dawson \cite{dawson}. Here we will focus on the electrostatic interaction
between a multistream quantum plasma described within the Schr\"odiner
model, a system first investigated in Ref.\ \cite{haas-etal1} (where also
the stationary regime was probed). Thus, we have the governing equations 
(\ref{eq:schrod-cont}) and (\ref{eq:schrod-mom}) but for $N$ beams of
electrons on a stationary ion background, \textit{i.e.}, Using this
decomposition in Eq.\ (\ref{eq:schrodinger}), we obtain 
\begin{equation}  \label{eq:schrod-cont-multi}
	\frac{d n_{\alpha}}{d t} = - n_\alpha \nabla \cdot \mathbf{v}_{\alpha} ,
\end{equation}
and 
\begin{equation}  \label{eq:schrod-mom-multi}
	\frac{d\mathbf{v}_{\alpha}}{d t} = \frac{e}{m_e} \nabla\phi 
	+ \frac{\hbar^2}{2m_e^2} \nabla 
	\left( \frac{ \nabla^2 \sqrt{ n_{\alpha} } }{ \sqrt{ n_{ \alpha } } } \right) ,
\end{equation}
now coupled through the self-consistent electrostatic potential governed by 
\begin{equation}
\nabla^2\phi = \frac{e}{\epsilon_0}\sum_{\alpha = 1}^N(n_{\alpha} - n_0).
\end{equation}
Here $n_0$ is the density of the stationary ion background.

In the one-stream case ($\alpha = 1$), we have the equilibrium solution $%
\mathbf{v} = \mathbf{v}_0$ (a constant drift relative the stationary ion
background) and the constant electron density $n = n_0$ (such that $\phi = 0$%
). Perturbing this system a Fourier decomposing the perturbations, such that 
$n = n_0 + \delta n\exp[i(\mathbf{k}\cdot\mathbf{x} - \omega t)])$, $\mathbf{%
v} = \mathbf{v}_0 + \delta\mathbf{v}\exp[i(\mathbf{k}\cdot\mathbf{x} -
\omega t)]$, and $\phi = \delta\phi\exp[i(\mathbf{k}\cdot\mathbf{x} - \omega
t)]$, we obtain \cite{Pines,haas-etal1} 
\begin{equation}
(\omega - \mathbf{k}\cdot\mathbf{v}_0)^2 = \omega_p^2 + \frac{\hbar^2k^4}{%
4m_e^2} ,
\end{equation}
where the last term is the Bohm--de Broglie correction to the dispersion
relation. Here we have the electron plasma frequency $\omega_p =
(e^2n_0/\epsilon_0m_e)^{1/2}$.

Similarly to the one-stream case, we obtain the dispersion relation \cite%
{haas-etal1,anderson-etal} 
\begin{eqnarray}
1 &=& \frac{\omega_{p1}^2}{(\omega - \mathbf{v}_{01}\cdot\mathbf{k})^2 -
\hbar^2k^4/4m_e^2}  \notag \\
&&\qquad + \frac{\omega_{p2}^2}{(\omega - \mathbf{v}_{02}\cdot\mathbf{k})^2
- \hbar^2k^4/4m_e^2} \ ,  \label{eq:twostreamfluid}
\end{eqnarray}
for two propagating electron beams (with velocities $\mathbf{v}_{01}$ and $%
\mathbf{v}_{02}$) with background densities $n_{01}$ and $n_{02}$. The
quantum effect has a subtle influence on the stability of the perturbed
plasma. For the case $n_{01} = n_{02} = n_0/2$ and $\mathbf{v}_{01} = -%
\mathbf{v}_{02} = \mathbf{v}_0$, we have the instability condition 
\begin{equation}
\frac{4}{{K}^2}\left( 1 - \frac{1}{{K}^2} \right) < H^2 < \frac{4}{{K}^2} ,
\end{equation}
in terms of the normalized wavenumber $K = kv_0/\omega_p$ and the quantum
parameter $H = \hbar\omega_p/m_ev_0^2$ (see Fig.\ 1) \cite%
{haas-etal1,anderson-etal}. We see that when $H = 0$, we have unstable
perturbations for $0 < K < 1$, but when $H \neq 0$ a considerably more
complex instability region develops.

A model for treating partial coherence in such systems, based on the Wigner
transform technique \cite{wigner,moyal,mendonca,schleich}, can also be
developed \cite{anderson-etal} (see also Ref.\ \cite{marklund}). Moreover,
using the equations (\ref{eq:pauli-cont}) and (\ref{eq:pauli-mom}), a
similar framework may be set up for electron streams with spin properties.


\subsection{Fluid model}



\subsubsection{Plasmas based on the Schr\"odinger model}

Suppose that we have $N$ electron wavefunctions, and that the total system
wave function can be described by the factorization 
$\psi(\mathbf{x}_1, \mathbf{x}_2, \ldots \mathbf{x}_N) = \psi_{1}\psi_{2} \ldots \psi_{N}$. 
For each wave function $\psi_\alpha$, we have a corresponding probability 
$\mathcal{P}_\alpha$. From this, we first define $\psi_\alpha =
n_\alpha\exp(iS_\alpha/\hbar)$ and follow the steps leading to Eqs.\ 
(\ref{eq:schrod-cont}) and (\ref{eq:schrod-mom}). We now have $N$ such equations
the wave functions $\{\psi_\alpha\}$. Defining \cite{Manfredi2005} 
\begin{equation}  \label{eq:meandensity}
n \equiv \sum_{\alpha = 1}^N \mathcal{P}_\alpha n_\alpha
\end{equation}
and 
\begin{equation}  \label{eq:meanvelocity}
\mathbf{v} \equiv \langle \mathbf{v}_\alpha \rangle = \sum_{\alpha = 1}^N 
\frac{\mathcal{P}_\alpha n_\alpha \mathbf{v}_\alpha}{n} ,
\end{equation}
we can define the deviation from the mean flow according to 
\begin{equation}
\mathbf{w}_\alpha = \mathbf{v}_\alpha - \mathbf{v} .
\end{equation}
Taking the average, as defined by (\ref{eq:meanvelocity}), of Eqs. (\ref%
{eq:schrod-cont}) and (\ref{eq:schrod-mom}) and using the above variables,
we obtain the quantum fluid equation 
\begin{equation}  \label{eq:cont}
	\frac{d n}{d t} = - n \nabla \cdot \mathbf v
\end{equation}
and 
\begin{equation}  \label{eq:mom-schrod}
	\frac{d \mathbf{v}}{dt} 
	= \frac{e}{m_e} \nabla \phi 
	- \frac{1}{n m_e} \nabla p + \frac{\hbar^2}{2m_e^2}\nabla 
	\left \langle \left(  \frac{ \nabla^2 \sqrt{n_\alpha} }{ \sqrt{n_\alpha} }\right) \right
	\rangle ,
\end{equation}
where we have assumed that the average produces an isotropic pressure $p =
m_en\langle |\mathbf{w}_\alpha|^2\rangle$ We note that the above equations
still contain an explicit sum over the electron wave functions. For typical
scale lengths larger than the Fermi wavelength $\lambda_F $, we may
approximate the last term by the Bohm--de Broglie potential \cite%
{Manfredi2005} 
\begin{equation}
\left\langle \frac{\nabla^2\sqrt{n_\alpha}}{\sqrt{n_\alpha}} \right\rangle
\approx \frac{\nabla^2\sqrt{n}}{\sqrt{n}} .
\end{equation}
Using a classical or quantum model for the pressure term, we finally have a
quantum fluid system of equations. For a self-consistent potential $\phi$ we
furthermore have 
\begin{equation}
\nabla^2\phi = \frac{e}{\epsilon_0}(n - n_i) .
\end{equation}


\subsubsection{Spin plasmas}


The collective dynamics of electrons with spin and some of the spin
modifications of the classical dispersion relation was presented in Ref.\ 
\cite{marklund-brodin}. Here we will follow Refs.\ \cite{marklund-brodin}
and \cite{brodin-marklund} for the derivation of the governing equations.
Suppose that we have $N$ wave functions for the electrons with magnetic
moment $\mu_e = -\mu_B$, and that, as in the case of the Schr\"odinger
description, the total system wave function can be described by the
factorization $\psi = \psi_{1}\psi_{2} \ldots \psi_{N}$. Then the density is
defined as in Eq.\ (\ref{eq:meandensity}) and the average fluid velocity
defined by (\ref{eq:meanvelocity}). However, we now have one further fluid
variable, the spin vector, and accordingly we let $\mathbf{S} = \langle%
\mathbf{s}_\alpha\rangle $. From this we can define the microscopic
microscopic spin density $\bm{\mathcal{S}}_{\alpha} = \mathbf{s}_{\alpha} - 
\mathbf{S}$, such that $\langle\bm{\mathcal{S}}_{\alpha} \rangle = 0$.

Taking the ensemble average of Eqs.\ (\ref{eq:pauli-cont}) we obtain the
continuity equation (\ref{eq:cont}), while we the the ensemble average
applied to (\ref{eq:pauli-mom}) yield 
\begin{equation}
	\frac{d v_i }{dt}  
	=	 	
	-  \frac{e}{m_e} \left(E_i + \epsilon_{ijk} v_j B_k \right) 
	- \frac{1}{n m_e} \frac{\partial p}{\partial x_i}   
	+ \frac{ \hbar^2 }{ 2m_e^2 } \frac{\partial}{\partial x_i}  
	\left( \frac{ \nabla^2 \sqrt{ n } }{ \sqrt{ n } } \right) 
	+ \frac{1}{n m_e} F^{\text{spin}}_i  
	\label{eq:mom-pauli}
\end{equation}
and the average of Eq.\ (\ref{eq:pauli-spin}) gives 
\begin{equation}
	\frac{d S_i}{d t}
	= 
	\frac{2\mu_B}{\hbar} \epsilon_{ijk} B_j S_k 
	- \frac{1}{n m_e} \frac{ \partial \Sigma_{ij}}{\partial x_j} 
	+ \frac{1}{n m_e} \Omega^{\text{spin}}_i  
	\label{eq:spin}
\end{equation}
respectively. Here the force density due to the electron spin is 
\begin{eqnarray}
	F^{\text{spin}}_i &=& 
	- \mu_B n   S_j \frac{\partial B_j}{\partial x_i} 
	- \frac{\hbar^2}{4m_e} \frac{\partial}{\partial x_j} 
	\left[ n \left( \Gamma_{ij}  + \tilde\Gamma_{ij} \right) \right]
	\nonumber \\ &&
	-\frac{\hbar^2}{4m_e} \frac{\partial}{\partial x_j} 
	\left[ n \left( \frac{\partial S_k}{\partial x_i} \right)	
	\left\langle \frac{\mathcal{S}_{\alpha k}}{\partial x_j} \right\rangle
	+ n \left\langle \frac{\partial \mathcal{S}_{\alpha k}}{\partial x_i} 
	\right\rangle   \frac{\partial S_{k}}{\partial x_j}  \right],
\end{eqnarray}
consistent with the results in Ref.\ \cite{degroot-suttorp}, while the
asymmetric thermal-spin coupling is 
\begin{equation}\label{K-tensor}
	\Sigma_{ij} = n m_e \langle \mathcal{S}_{\alpha i} w_{\alpha j}  \rangle
\end{equation}
and the nonlinear spin fluid correction is 
\begin{eqnarray}
	\Omega^{\text{spin}}_i 
	&=& 
	\frac{\hbar}{2} \epsilon_{ijk} S_j \left[ 
	\frac{\partial}{\partial x_a} \left( n \frac{ \partial S_k}{\partial x_a}  \right) \right]
	+ \frac{\hbar}{2}  \epsilon_{ijk} S_j 
	\left[ \frac{\partial}{\partial x_a}  
	\left( n \left< \frac{\partial \mathcal{S}_{\alpha k}}{\partial x_a} \right> \right) \right]  
	\nonumber \\ &&
	+ \frac{n \hbar}{2}  \epsilon_{ijk} \left\langle 
	\frac{\mathcal{S}_{\alpha j}}{n_{\alpha}} 
	\left\{ \frac{\partial}{\partial x_a} \left[ n_{\alpha} \frac{\partial}{\partial x_a} 
	( S_k + \mathcal{S}_{\alpha k}) \right] \right\} 
	\right\rangle 
	\label{Omega}
\end{eqnarray}
where $\Gamma_{ij} = ( \partial_i S_a) (\partial_j S_a )$ is the nonlinear spin 
correction to the classical momentum equation, $\tilde{\Gamma}_{ij} 
= \langle (\partial_i \mathcal{S}_{(\alpha )a}) (\partial_j \mathcal{S}_{(\alpha )}^{a})\rangle $
is a pressure like spin term (which may be decomposed into trace-free part
and trace). We note that, apart from the additional spin density evolution equation 
(\ref{eq:spin}), the momentum conservation equation (\ref{eq:mom-pauli}) is considerably more
complicated compared to the Schr\"odinger case represented by (\ref%
{eq:mom-schrod}). Moreover, Eqs.\ (\ref{eq:mom-pauli}) and (\ref{eq:spin})
still contains the explicit sum over the $N$ states, and has to be
approximated using insights from quantum kinetic theory or some effective
theory.

The coupling between the quantum plasma species is mediated by the
electromagnetic field. By definition, we let $\mathbf{H} = \mathbf{B}/\mu_0
- \mathbf{M}$ where $\mathbf{M} = -2n\mu_B\mathbf{S}/\hbar$ is the
magnetization due to the spin sources. Amp\`ere's law $\mathbf{\nabla}\times%
\mathbf{H} = \mathbf{j} + \epsilon_0\partial_t\mathbf{E}$ takes the form 
\begin{equation}
\mathbf{\nabla} \times \mathbf{B}=\mu _{0}(\mathbf{j} + \mathbf{\nabla}
\times \mathbf{M}) + \frac{1}{c^2}\frac{\partial\mathbf{E}}{\partial t},
\label{Eq-ampere}
\end{equation}
where $\mathbf{j}$ is the free current contribution The system is closed by
Faraday's law 
\begin{equation}
\mathbf{\nabla} \times \mathbf{E}=-\frac{\partial\mathbf{B}}{\partial t} .
\label{Eq-Faraday}
\end{equation}


\subsection{The magnetohydrodynamic limit}


The concept of a magnetoplasma was first introduced in the pioneering work 
\cite{alfven} by Alfv\'en, who showed the existence of waves in magnetized
plasmas. Since then, magnetohydrodynamics (MHD) has found applications in a
vast range of fields, from solar physics and astrophysical dynamos, to
fusion plasmas and dusty laboratory plasmas.

Magnetic fields, an essential component in the MHD description of plasmas,
also couples directly to the spin of the electron. Thus, the presence of
spin alters the single electron dynamics, introducing a correction to the
Lorentz force term. Indeed, from the experimental perspective, a certain
interest has been directed towards the relation of spin properties to the
classical theory of motion (see, e.g., Refs.\ \cite%
{halperin-hohenberg,balatsky,rathe-etal,hu-keitel,arvieu-etal,aldana-roso,walser-keitel,qian-vignale,walser-etal,roman-etal,liboff,fuchs-etal,kirsebom-etal}%
). In particular, the effects of strong fields on single particles with spin
has attracted experimental interest in the laser community \cite%
{rathe-etal,hu-keitel,arvieu-etal,aldana-roso,walser-keitel,walser-etal}.
However, the main objective of these studies was single particle dynamics,
relevant for dilute laboratory systems, whereas our focus will be on
collective effects.

We will now include if the ion species, which are assumed to be described by
the classical equations and have charge $Ze$, we may derive a set of one-
fluid equations \cite{brodin-marklund}. The ion equations read 
\begin{equation}
	\frac{d n_{I}}{d t} 
	= -  n_I \nabla \cdot \mathbf{v}_{I} ,  
	\label{eq:ion-density}
\end{equation}
and 
\begin{equation}
	m_{I} n_{I} \frac{d v_{I i}}{dt} 
	=
	Zen_{I}\left( E_i
	+ \epsilon_{ijk} v_{Ij}  B_k \right) 
	- \frac{ \partial p_{I}}{\partial x_i}  ,
	\label{eq:ion-mom}
\end{equation}
Next we define the total mass density $\rho \equiv (m_{e}n + m_{I}n_ {I})$,
the centre-of-mass fluid flow velocity $\mathbf{V}\equiv (m_{e}n \mathbf{v}%
_{e}+m_{I}n_{I}\mathbf{v}_{I})/\rho $, and the current density $\mathbf{j}=-
en\mathbf{v}_{e}+Zen_{I}\mathbf{v}_{I}$. Using these denfinitions, we
immediately obtain 
\begin{equation}
\frac{d \rho}{d t}
= -  \rho \nabla \cdot \mathbf{V},
\label{eq:mhd-cont}
\end{equation}
from Eqs.\ (\ref{eq:cont}) and (\ref{eq:ion-density}). Assuming
quasi-neutrality, i.e.\ $n \approx Zn_{I}$, the momentum conservation
equations (\ref{eq:mom-pauli}) and (\ref{eq:ion-mom}) give 
\begin{equation}
	\rho \frac{ d V_i}{dt} = \epsilon_{ijk} j_j B_k 
	- \frac{\partial \mathsf{\Pi}_{ij}}{\partial x_j}
	- \frac{\partial p}{\partial x_i} 
	+ \frac{Z\hbar^2\rho}{2m_em_I}
	\frac{\partial}{\partial x_i} \left( \frac{\nabla^2\sqrt{\rho}}{\sqrt{\rho}}\right) 
	+ F^{\text{spin}}_i,  
	\label{eq:mhd-mom}
\end{equation}
where $\mathbf{\mathsf{\Pi}}$ is the tracefree pressure tensor in the
centre-of-mass frame, and $P$ is the scalar pressure in the centre-of-mass
frame. We also note that due to quasi-neutrality, we have $n_ {e} \approx
Z\rho /m_{I}$ and $\mathbf{v} = \mathbf{V} - m_{I}\mathbf{j}/Ze\rho $, and
we can thus express the quantum terms in terms of the total mass density $%
\rho $, the centre-of-mass fluid velocity $\mathbf{V}$, and the current $%
\mathbf{j}$. With this, the spin transport equation (\ref{eq:spin}) reads 
\begin{equation}
	\rho \frac{d S_i}{dt} 
	= \frac{m_{e}}{Ze} j_j \frac{\partial S_i}{\partial x_j} 
	+ \frac{2\mu_B \rho } {\hbar } \epsilon_{ijk} B_j S_k 
	- \frac{m_{I}}{Z} \frac{ \partial \Sigma_{ij} }{\partial x_j} 
	+ \frac{m_{I}}{Z} \Omega^{\text{spin}}.  
	\label{eq:mhd-spin}
\end{equation}

In the momentum equation (\ref{eq:mhd-mom}), neglecting the pressure and the
Bohm--de Broglie potential for the sake of clarity, we have the force
density $\mathbf{j}\times\mathbf{B} + \mathbf{F}_{\text{spin}}$. In general,
for a magnetized medium with magnetization density $\mathbf{M}$, Amp\`ere's
law gives the free current in a finite volume $V$ according to 
\begin{equation}  \label{eq:free-current}
\mathbf{j} = \frac{1}{\mu_0}\mathbf{\nabla}\times\mathbf{B} - \mathbf{\nabla}%
\times\mathbf{M} ,
\end{equation}
where we have neglected the displacement current. The surface current is an
important part of the total current when we are interested in the forces on
a finite volume, as was demonstrated in Ref.\ \cite{brodin-marklund} and
will be shown below.

It it worth noting that the expression of the force density in the momentum
conservation equation can, to lowest order in the spin, be derived on
general macroscopic grounds. Formally, the total force density on a volume
element $V$ is defined as $\mathbf{F} = \lim_{V \rightarrow 0}(\sum_{\alpha} 
\mathbf{f}_{\alpha}/V)$, where $\mathbf{f}_{\alpha}$ are the different
forces acting on the volume element, and might include surface forces as
well. For magnetized matter, the total force on an element of volume $V$ is
then 
\begin{equation}
\mathbf{f}_{\mathrm{tot}} = \int_V\mathbf{j}_{\mathrm{tot}}\times\mathbf{B}\,%
\mathrm{d}V + \oint_{\partial V}(\mathbf{M}\times\hat{\mathbf{n}})\times%
\mathbf{B}\,\mathrm{d}S
\end{equation}
where (neglecting the displacement current) $\mathbf{j}_{\mathrm{tot}} = 
\mathbf{j} + \mathbf{\nabla}\times\mathbf{M}$. Inserting the expression for
the total current into the volume integral and using the divergence theorem
on the surface integral, we obtain the force density 
\begin{equation}
\mathbf{F}_{\mathrm{tot}} = \mathbf{j}\times\mathbf{B} + M_k\mathbf{\nabla}%
B^k ,
\end{equation}
identical to the lowest order description from the Pauli equation (see Eq.\ (%
\ref{eq:mhd-mom})). Inserting the free current expression (\ref%
{eq:free-current}), due to Amp\`ere's law, we can write the total force
density according to 
\begin{equation}  \label{eq:total-force}
{F}^i = -\partial^i\left( \frac{B^2}{2\mu_0} - \mathbf{M}\cdot\mathbf{B}
\right) + \partial_k({H^iB^k}) .
\end{equation}
The first gradient term in Eq.\ (\ref{eq:total-force}) can be interpreted as
the force due to a potential (the energy of the magnetic field and the
magnetization vector in that field), while the second divergence term is the
anisotropic magnetic pressure effect. Noting that the spatial part of the
stress tensor takes the form \cite{degroot-suttorp} 
\begin{equation}
T^{ik} = -H^iB^k +(B^2/2\mu_0 - \mathbf{M}\cdot\mathbf{B} )\delta^{ik} ,
\end{equation}
we see that the total force density on the magnetized fluid element can be
written $F^i = -\partial_kT^{ik}$, as expected. Thus, the Pauli theory
results in the same type of conservation laws as the macroscopic theory. The
momentum conservation equation (\ref{eq:mhd-mom}) then reads 
\begin{equation}
\rho \left( \frac{\partial }{\partial t}+\mathbf{V}\cdot \mathbf{\nabla}
\right)\mathbf{V} = -\mathbf{\nabla}\left( \frac{B^2}{2\mu_0} - \mathbf{M}%
\cdot\mathbf{B} \right) + B^k\partial_k\mathbf{H} -\mathbf{\nabla}p ,
\label{eq:mhd-mom2}
\end{equation}
where for the sake of clarity we have assumed an isotropic pressure, dropped
the displacement current term in accordance with the nonrelativistic
assumption, and neglected the Bohm potential (these terms can of course
simply be added to (\ref{eq:mhd-mom2})). This concludes the discussion of
the spin-MHD plasma case. Next, we will look at some applications of the
derived equations. However, it should be noted that in many cases the spins
are close to thermodynamic equilibrium, and we can thus write the
paramagnetic electron response in terms of the magnetization \cite%
{brodin-marklund} 
\begin{equation}
\mathbf{M}=\frac{\mu _{B}\rho}{m_I}\,\tanh \left( \frac{\mu _{B}B}{k_BT}%
\right) \widehat{\mathbf{B}} ,  \label{Eq: tanh-factor}
\end{equation}
instead of using the full spin dynamics. Here $B$ denotes the magnitude of
the magnetic field and $\widehat{\mathbf{B}}$ is a unit vector in the
direction of the magnetic field, $k_B$ is Boltzmann's constant, and $T$ is
the electron temperature.


\section{Spin Quantum Kinetics}

%
Quantum mechanics in terms of quasi-distribution functions is perhaps the
formulation with the closest resemblance to classical statistical mechanics.
The formulation started with Wigner's paper \cite{wigner} together with
Weyls correspondence principle \cite{weyl}. In terms of this formulation the
state of the system is no longer described by a density matrix but instead a
phase-space distribution function. Similarly operators are translated into
phase functions. Calculating the expectation value of an operator is then a
matter of calculating a phase space integral a corresponding function
weighted by the distribution function. The method has been applied to a wide
range of problems. For example it has been applied in optics \cite%
{glauber2,leonhardt}, collision theory \cite{lee2,carruthers}, nonlinear
theory \cite{takahashi} and transport problems in solid state physics, see
for example \cite{haug,rammer} and references therein.

There are many different ways to define a quasi distribution function in
quantum mechanics. The most known examples are probably the
Glauber-Sudarshan p-distribution \cite{glauber, sudarshan}, the
q-distribution \cite{glauber65} and the related Husimi distribution \cite%
{husimi}. The many different definitions basically comes from the fact that
the position and momentum operators do not commute, so the transformation
between an operator and a phase space function is not unique. See for
example \cite{lee} for a review of the different phase space distribution
functions. The Wigner distribution corresponds to ordering the position and 
\textit{canonical} momentum operators symmetrically.

When considering a particle in a magnetic field gauge invariance has to be
assured. This can be done by adding a phase factor to the definition of the
Wigner function \cite{stratonovich}. The phase factor will then compensate
for the change of phase that occurs in the density matrix when performing a
gauge transformation. When dealing with the gauge invariant
Wigner-Stratonovich the Weyl correspondence is modified \cite{serimaa}. The
natural variables to use are the position and \textit{kinetic} momentum.

The formulation of quantum mechanics in terms of phase space distributions
has also been generalized for spin particles \cite%
{cohen,chandler,scully,cunha}. Also in this case the definition is not
unique and one can find analogs to the different definitions in the phase
space case \cite{cohen}.

\subsection{Scalar quasi-distribution theory for a spin plasma}

The distribution function which we will work with here is the combination of
the Wigner distribution function \cite{wigner} for the phase-space variables
and the q-function for the spin degree of freedom \cite{scully}. This
combination of distribution functions was used in \cite{zamanian2010} and it
turns out to yield a intuitive description of spin-1/2 particles in an
extended phase space.

Given a 2-by-2 density matrix (in the Schr\"odinger pricture $\rho (\mathbf{x},%
\mathbf{y},t)$ for a spin particle the extended phase-space distribution
function is defined by 
\begin{equation}
f(\mathbf{x},\mathbf{p},\mathbf{s},t)=\frac{1}{4\pi }\mathrm{Tr}\left[
\left( 1+\mathbf{s}\cdot \bm{\sigma}\right) W(\mathbf{x},\mathbf{p},t)\right]
,  \label{j1}
\end{equation}%
where $\mathrm{Tr}$ denotes that the trace is to be calculated of the
resulting 2-by-2 matrix and where $\bm\sigma $ is a vector with the Pauli
matrices as components, $\mathbf{s}$ is a vector on the unit sphere. The
Wigner distribution matrix function is given by 
\begin{equation}
W(\mathbf{x},\mathbf{p},t)=\int \frac{d^{3}y}{(2\pi \hbar )^{3}}e^{-\frac{i}{%
\hbar }\mathbf{y}\cdot \left[ \mathbf{p}+iq\int_{-1/2}^{1/2}ds\mathbf{A}(%
\mathbf{x}+s\mathbf{y},t)\right] }\rho \left( \mathbf{x}+\frac{\mathbf{y}}{2}%
;\mathbf{x}-\frac{\mathbf{y}}{2},t\right) ,  \label{j2}
\end{equation}%
where $q$ is the charge of the particle and $\mathbf{A}$ is the vector
potential. The integral over the vector potential is there to ensure gauge
invariance. The momentum variable $\mathbf{p}$ is the gauge invariant
kinetic momentum related to the canonical momentum $\mathbf{p}_{c}$ by $%
\mathbf{p}=\mathbf{p}_{c}-q\mathbf{A}(\mathbf{x},t)$. This distribution
function is defined on an extended phase-space $(\mathbf{x},\mathbf{p},%
\mathbf{s})$ and can in principle be used to calculate the expectation value
of any observable defined by an operator $\hat{O}=O(\hat{\mathbf{x}},\hat{%
\mathbf{p}},\bm\sigma )$. The way to do this is to use the modified Weyl
correspondence \cite{serimaa} together with a transformation for the spin
variable to obtain a phase-space function, and subsequently take the average
of this function weighted by the distribution function. For an operator
depending on $\hat{\mathbf{x}},\hat{\mathbf{p}}$ the corresponding
phase-space function is obtained by 
\begin{equation}
O(\mathbf{x},\mathbf{p})=\int d^{3}ye^{-\frac{i}{\hbar }\mathbf{y}\cdot %
\left[ \mathbf{p}+q\int_{-1/2}^{1/2}ds\mathbf{A}(\mathbf{x}+s\mathbf{y},t)%
\right] }\left\langle \mathbf{x}+\frac{\mathbf{y}}{2}\right\vert \hat{O}%
\left\vert \mathbf{x}-\frac{\mathbf{y}}{2}\right\rangle 
\end{equation}%
It can sometimes be found more easily by first putting the position and
kinetic momentum operators of the operator $\hat{O}$ in symmetric order
using the commutation relation and then make the substitution $\hat{\mathbf{x%
}}\rightarrow \mathbf{x}$ and $\hat{\mathbf{p}}\rightarrow \mathbf{p}$. For
example the pressure tensor which just contains the kinetic momentum
operator is obtained by $\hat{P}_{ij}=\hat{p}_{i}\hat{p}_{j}\rightarrow
p_{i}p_{j}$. Note that the gauge dependent Wigner distribution this
correspondence is not so simple anymore, since we are then dealing with the
position and canonical momentum operators which has to be ordered
symmetrically using the commutation relation. So for the moment above, for
example we have to put the combination $\hat{P}_{ij}=[\hat{p}_{ci}-qA_{i}(%
\hat{\mathbf{x}})][\hat{p}_{cj}-qA_{j}(\hat{\mathbf{x}})]$ in symmetric
ordering and then make the substitution $\hat{x}_{i}\rightarrow x_{i}$ and $%
\hat{p}_{i}\rightarrow p_{i}$. This is a quite difficult task for a general
operator since the form of the vector potential is not necessarily known.

The spin space function corresponding an operator $\hat O$ depending on $\bm %
\sigma$ is obtained by transformation 
\begin{equation}
O (\mathbf{s}) = \frac{ 1}{4 \pi} \mathrm{Tr} \left[ \left( 1 + \mathbf{s }%
\cdot \bm \sigma \right) \hat O\right] .  \label{spintrans}
\end{equation}
If the operator in question depends on both the position and momentum
operators and the spin, both of these transformations have to be made, see
Ref.\ \cite{}. Since $\bm \sigma ^2 = 1$ the only possible spin operators we
may have is the identity operator 1 and $\bm \sigma$. The transformation %
\eqref{spintrans} above yields 
\begin{equation}
\hat O = \bm \sigma \rightarrow O(\mathbf{s}) = 3 \mathbf{s}.  \label{j5}
\end{equation}

The momentum variable in the Wigner function above is the canonical
momentum. In the presence of a magnetic field it is often more convenient to
work with the gauge invariant kinetic momentum.

\subsection{Evolution and the long scale length limit}

The Hamiltonian for a spin-1/2 particle in a magnetic field is given by 
\begin{equation}
\hat{H}=\frac{[\hat{\mathbf{p}}_{c}-A(\hat{\mathbf{x}},t)]^{2}}{2m}+qV+\mu
_{B}\bm\sigma \cdot \mathbf{B}(\hat{\mathbf{x}},t),
\end{equation}%
where $V$ and $\mathbf{A}$ are the electromagnetic potentials, $\mathbf{B}%
=\nabla \times \mathbf{A}$ is the magnetic induction and $\mu _{B}$ is the
magnetic moment of the particle. Specifically, for an electron, the magnetic
moment is given by $\mu _{B}=qg\hbar /(4m)$ where $g\approx 2.001$ is a
correction factor deduced from quantum electrodynamics \cite{schwinger}.
Note that we have used $q=-|e|$ so that the magnetic moment is negative. The
evolution equation for the density operator is given by the von Neumann
equation 
\begin{equation}
i\hbar \partial _{t}\rho =[\rho ,\hat{H}].
\end{equation}%
Taking the transform \eqref{j1} of this equation, it is possible to derive
an evolution equation for the extended phase space Wigner function $f(%
\mathbf{x},\mathbf{p},\mathbf{s},t)$, see Ref.\ \cite{zamanian2010} and it
is given by 
\begin{eqnarray}
\frac{\partial f}{\partial t}+(\mathbf{v}+\Delta \tilde{\mathbf{v}})\cdot
\nabla _{x}f+\frac{q}{m}\left[ \tilde{\mathbf{E}}+(\mathbf{v}+\tilde{\mathbf{%
v}})\times \tilde{\mathbf{B}})\right] \cdot \nabla _{v}f &&  \notag \\
+\frac{\mu _{B}}{m}\nabla _{x}\left[ (\mathbf{s}+\nabla _{s})\cdot \tilde{%
\mathbf{B}}\right] \cdot \nabla _{v}f+\frac{2\mu _{B}}{\hbar }(\mathbf{s}%
\times \tilde{\mathbf{B}})\cdot \nabla _{s}f &=&0.  \label{j8}
\end{eqnarray}%
where $\mathbf{v}=\mathbf{p}/m$ is the velocity and we have defined the
operators 
\begin{eqnarray}
\tilde{\mathbf{E}} &=&\mathbf{E}(\mathbf{x})\int_{-1/2}^{1/2}ds\cos \left( 
\frac{\hbar s}{m}\overleftarrow{\nabla }_{x}\cdot \overrightarrow{\nabla }%
_{v}\right)  \\
\tilde{\mathbf{B}} &=&\mathbf{B}(\mathbf{x})\int_{-1/2}^{1/2}ds\cos \left( 
\frac{\hbar s}{m}\overleftarrow{\nabla }_{x}\cdot \overrightarrow{\nabla }%
_{v}\right)  \\
\Delta \tilde{\mathbf{v}} &=&\frac{q\hbar }{m^{2}}\mathbf{B}(\mathbf{x}%
)\int_{-1/2}^{1/2}dss\sin \left( \frac{\hbar s}{m}\overleftarrow{\nabla }%
_{x}\cdot \overrightarrow{\nabla }_{v}\right) \cdot \nabla _{v}.
\end{eqnarray}%
Note the similarity of the equation above with the classical Vlasov
equation. In order to compare it further we may consider the semiclassical
limit which is applicable when the typical length scale is much shorter than
the de Broglie wave length $\hbar /(mv)$, where $v$ is the typical velocity
of the system. We may then expand the sine and cosine operators above.
Keeping terms up to order $\hbar ^{2}$ the resulting equation is 
\begin{equation}
\frac{\partial f}{\partial t}+\mathbf{v}\cdot \nabla _{x}f+\frac{q}{m}\left[ 
\mathbf{E}+\mathbf{v}\times \mathbf{B}\right] +\frac{\mu _{B}}{m}\nabla _{x}%
\left[ (\mathbf{s}+\nabla _{s})\cdot \mathbf{B}\right] \cdot \nabla _{v}f+%
\frac{2\mu _{B}}{\hbar }(\mathbf{s}\times \mathbf{B})\cdot \nabla _{s}f=0.
\label{j12}
\end{equation}%
The first term proportional to the spin $\mathbf{s}$ is the dipole force of
a magnetic moment in an inhomogeneous magnetic field and the last term is
accounts for the spin precession. Both of these can be understood from a
classical analog, however, there is also a \textit{quantum dipole} term
given by $(\mu _{B}/m)\nabla _{x}(\mathbf{B}\cdot \nabla _{s})\cdot \nabla
_{v}f$. This term accounts for the fact that the spin is not a classical
dipole and we can for example not have a distribution function proportional
to $\delta (\mathbf{s}-\hat{z})$, where $\hat{z}$ is a unit vector in the
z-direction. The corresponding distribution function in the quantum
mechanical case is proportional to $1+\cos \theta _{s}$ where $\theta _{s}$
is the angle between the spin direction and the z-axis. The macroscopic
magnetization the classical and quantum cases are still the same which is
ensured by the factor 3 in the transformation \eqref{j5} above. This factor
occurs in the quantum case 
\begin{equation}
\mathbf{M}_{\text{q}}(\mathbf{x})=3\mu _{B}\int d\Omega f(\mathbf{x},\mathbf{%
v},\mathbf{s}),
\end{equation}%
but is absent in the corresponding classical equation 
\begin{equation}
\mathbf{M}_{\text{cl}}(\mathbf{x})=\mu _{B}\int d\Omega \mathbf{s}f_{\text{cl%
}}(\mathbf{x},\mathbf{v},\mathbf{s}),
\end{equation}%
where $f_{\text{cl}}$ is the classical distribution function and we in both
cases have $d\Omega =d^{3}vd^{2}s$.

A semi classical version of Eq.\ \eqref{j12} (where the quantum dipole term
is missing) has been applied to find a new type of resonance due to
radiative corrections to the electron $g$-factor \cite{brodin2008}. A
related kinetic equation has also been applied in the context of
thermonuclear fusion \cite{cowley}. In this model, however, both the dipole
and the quantum dipole terms are missing. This equation can formally be
retained from our result above by neglecting all terms of order $\hbar$ or
higher.

\subsection{Thermal equilibrium}

The scalar spin distribution does also have some other important differences
compared to the distribution for a classical dipole. An important example is
the thermal equilibrium distribution. For in the classical case it would be
given by 
\begin{equation}
f_{\mathrm{cl}}^T (\mathbf{v }, \mathbf{s}) = n_0 f_M(v) \frac{1}{ 4 \pi } 
\frac{\mu_B B_0}{k_B T} \left[ \sinh \left( \frac{\mu_B B_0}{k_B T} \right) %
\right]^{-1} \exp \left[ - \frac{\mu_B \mathbf{B}_0 \cdot \mathbf{s}}{k_B T} %
\right] ,
\end{equation}
where $n_0$ is the equilibrium density, $f_M$ is a normalized Maxwellian
velocity distribution, $k_B$ is Boltzmann's constant, $T$ is the temperature
and $\mathbf{B}_0$ is the external magnetic field. This distribution gives
rise to a magnetization 
\begin{equation}
\mathbf{M}_{\text{cl}}^T = n_0 \mu_B \eta \left( \frac{\mu_B B_0}{k_B T}
\right),
\end{equation}
where $\eta$ is the Langevin function. In the quantum mechanical case the
corresponding distribution function becomes (assuming that the chemical
potential is sufficiently large so that Landau quantization can be
neglected, see Ref.\ \cite{zamanian2010}) 
\begin{equation}
f_{\mathrm{q}}^T (\mathbf{v}, \mathbf{s}) = n_0 f_M (\mathbf{v}) \frac{1}{%
4\pi} \left[ 1 + \tanh \left( \frac{\mu_B B_0}{k_B T} \right) \cos \theta_s %
\right] ,
\end{equation}
where $f_M$ it the usual Maxwell-Boltzmann distribution. The zeroth order
magnetization in this case is given by 
\begin{equation}
\mathbf{M}_{\text{q}}^T = 3 \mu_B \int d\Omega \mathbf{s }f_{\mathrm{q}} =
n_0 \mu_B \tanh\left( \frac{\mu_B B_0}{k_B T} \right) ,
\end{equation}
as expected. The dynamics of the magnetization can be treated by the use of
a fluid moment hierarchy and we now go on to consider this.

\section{Spin Fluid Moments}

Many problems do not require the full machinery of the kinetic approach and
calculations can be greatly simplified if executed in a macroscopic fluid
model instead (see for example \cite{nicholson} and references therein). In
particular when dealing with nonlinear problems the kinetic approach soon
becomes very cumbersome [refs?] and the need for a simplified theory cannot
be understated.

In this section such a theory will be presented, derived from the kinetic
theory by taking moments of the quantum kinetic equation \cite{zamanian2010}
in a way analogous to what is done in classical plasma theory (see for
example \cite{nicholson}). The theory presented here was derived in \cite%
{zamanian2010b}. As in the classical approach all intrinsically kinetic
features such as Landau damping will be lost, and one is faced with a
closure problem, since the fluid hierarchy is an infinite series of
equations that needs to be truncated at some point. 

Quantum fluid moments derived from the Wigner formalism have been applied
before, see for example Refs.\ \cite{haas0,haas1}, and also in the case of
gauge invariant Wigner-Stratonovich formalism\cite{haas2}. In the spin-1/2
case the moments have also been calculated \cite{balescu} starting from a
quantum kinetic equation \cite{zhang}. In their treatment they use a matrix
form of the kinetic equation and they also retain the collision terms which
we here will neglect. However, their fluid hierarchy is only discussed
shortly and also it is derived to order $\hbar$ and hence, for example the
effect of the dipole term on the dynamics was not retained.

Since we are working with a quasi distribution function in a phase space
extended to also include the microscopic spin variable, the classical
approach must be slightly modified. Firstly we also need to integrate over
the microscopic spin variable, and furthermore a new macroscopic spin
variable is defined, leading to a new hierarchy of spin-dependent
macroscopic objects \cite{zamanian2010b}. Thus we define the moments as 
\begin{eqnarray}
n &=& \int d\Omega f ,  \label{density} \\
\mathbf{u }&=& \frac{ 1}{n} \int d\Omega \mathbf{v }f ,  \label{velocity} \\
P_{ij} &=& m \int d\Omega (v_i - u_i) (v_j - u_j) f ,  \label{pressure} \\
Q_{ijk} &=& m \int d\Omega (v_i - u_i) (v_j - u_j) (v_k - u_k) f
\label{energyflux} \\
\mathbf{S }&=& \frac{ 3}{n} \int d\Omega \mathbf{s }f ,  \label{spindens} \\
\Sigma_{ij} &=& m \int d\Omega (3 s_i - S_i) (v_j - u_j) f ,  \label{spinvel}
\\
\Lambda_{ijk} &=& m \int d\Omega (s_i - S_i) (v_j - u_j) (v_k - u_k) f .
\label{svv}
\end{eqnarray}
Here the first four moments are respectively the density, the fluid
velocity, the pressure density and the energy flux density. The equation
moment above \eqref{spindens} defines the spin density $\mathbf{S }= \mathbf{%
S }(\mathbf{x},t)$ which yields the average spin density at position $%
\mathbf{x}$ and time $t$. The factor 3 in this definition occurs due to the
correspondence \eqref{j5}. The sixth moment \eqref{spinvel} is a mixed
moment of the velocity and the spin which will act as some kind of
spin-pressure. Finally, we have a mixed spin-velocity-velocity moment which
could perhaps be termed the spin-pressure correlation. Similarly we could go
on to define even higher order moments like the energy flux or a higher
order mixed moment. Note that there is no need to include higher order
moments in the spin variable since we have that $\int d\Omega \hat s_i \hat
s_j \propto \delta_{ij}$.

Using the evolution equation for the extended phase space distribution
function \eqref{j12} we may now calculate the evolution equation of the
different moments 
\begin{eqnarray}
	\frac{dn}{dt} &=&-n\nabla \cdot \mathbf{u}, 
 	\label{cont-mom-exp} 
	\\
	\frac{du_{i}}{dt} &=& \frac{q}{m}(E_{i} + \epsilon _{ijk}u_{j}B_{k})
	+ \frac{\mu }{m}S_{j}\frac{\partial B_{j}}{\partial x_{i}}
	- \frac{1}{nm} \frac{\partial P_{ij}}{\partial x_{j}}, \\
	\frac{dP_{ij}}{dt} &=& - P_{ik}\frac{\partial U_{j}}{\partial x_{k}}
	- P_{jk} \frac{\partial U_{i}}{\partial x_{k}}
	- P_{ij}\frac{\partial u_{k}}{\partial x_{k}}
	+ \frac{q}{m}\epsilon _{imn} P_{jm}B_{n} 
	+ \frac{q}{m}\epsilon_{jmn}P_{im}B_{n} 
	+ \frac{\mu }{m}\Sigma _{ik}\frac{\partial B_{k}}{\partial x_{j}}  
	\notag \\
	&& 
	+ \frac{\mu }{m}\Sigma _{jk} \frac{\partial B_{k}}{\partial x_{i}}
	-\frac{\partial Q_{ijk}}{\partial x_{k}}, 
	\\
	\frac{dS_{i}}{dt} &=& \frac{2\mu }{\hbar }\epsilon_{ijk}S_{j}B_{k}
	- \frac{1}{nm}\frac{\partial \Sigma _{ij}}{\partial x_{j}} 
	\label{spindensev}
	\\
	\frac{d\Sigma _{ij}}{dt} &=&
	- \Sigma _{ij}\frac{\partial U_{k}}{\partial x_{k}} 
	- \Sigma _{ik}\frac{\partial U_{j}}{\partial x_{k}}
	- P_{jk}\frac{\partial S_{i}}{\partial x_{k}}
	+ \frac{q}{m}\epsilon _{jkl}\Sigma _{ik}B_{l}
	+ \frac{ 2\mu }{\hbar } \epsilon_{ikl} \Sigma_{kj}B_{l}
	+ \mu n\frac{\partial B_{i}}{\partial x_{j}}  
	\notag 
	\\
	&& 
	- \mu nS_{i}S_{k}\frac{\partial B_{k}}{\partial x_{j}}
	- \frac{\partial \Lambda _{ijk}}{\partial x_{k}},  
	\label{14}
\end{eqnarray}
where $d/dt=\partial _{t}+\mathbf{v}\cdot \nabla $. It is possible to derive
the evolution equation for the $Q_{ijk}$, see \cite{haas1} and $\Lambda
_{ijk}$, however, these moments will then couple to even higher order
moments. Instead some approximation is needed. In our paper \cite%
{zamanian2010b} we used the closure $Q_{ijk}=0$ and $\Lambda _{ijk}=0$,
which is perhaps the easiest one to deal with.

As can be seen in Eq.\ \eqref{spindensev}, the spin-velocity moment $%
\Sigma_{ij}$ acts as a pressure term in the evolution equation for the spin
density. By including this moment, it is possible to capture some kinetic
effects in a fluid theory. Some problems which might be difficult or tedious
to solve in a kinetic theory may be reachable within a fluid theory.
However, the exact role of the spin-velocity moment is a subject of further
research.

\subsection{Two-fluid model}

It is shown in the previous section treating the kinetic equation that the
distribution function can be divided in two parts, one for each spin
direction along the magnetic field. Each fluid is seen to obey the same
hierarchy as above \cite{zamanian2010b}, and we will just have two sets of
these equations, one for each species. Of course we will have separate
macroscopic quantities $n_\alpha$, $\bm v_\alpha$, $\bm S_\alpha$ and $%
\Sigma_\alpha$ for each species. Here the subscript $\alpha$ indicates which
species the quantities refer to. This approach adds a bit of complexity but
can capture some kinetic effects due to a the different dynamics of the two
spin states, and is therefore worth pursuing in problems were such physics
is expected to play a role\cite{brodinSpinPond}.


\section{Applications of fluid and kinetic models}


The various quantum models developed cover several physical effects. Effects
of the Fermi pressure and particle dispersion has been described in some
detail in Refs. \cite{Manfredi2005,Shukla-Eliasson} both within fluid
theories and within a kinetic approach. The fluid approach uses the so
called Bohm-de Broglie potential to get an effective quantum force in the
momentum equation, and the equation of state is chosen such as to get the
Fermi pressure. Several modifications of classical behavior due to such
models has been described in the literature, see the references in Ref. \cite%
{Shukla-Eliasson} for an up-to-date list. Many papers using kinetic
approaches cover the effects of particle dispersion and Fermi pressure, but
using a Wigner function derived without the magnetic dipole coupling of the
Pauli Hamiltonian. In this way all effects due to the magnetic dipole force
and spin magnetization is left out. In contrast to the these works, we will
here focus on physical effects directly associated with the spin-coupling in
the Pauli-Hamiltonian, which give raise to the magnetic dipole force, the
spin precession and the spin magnetization in the above presented models.
Most of the recent results along these lines has been derived from models
similar to those presented here.

\subsection{Results from spin-fluid theories}

The most basic question to ask concerning the electron spin-properties in a
plasma is "when are they important?". For spin effects due to the
Fermi-pressure this is straightforward to answer. The Fermi-pressure becomes
important when the Fermi temperature approaches the thermodynamic
temperature, which give a simple condition on the temperature and density of
a plasma. For the effects due to the direct spin-coupling in the
Pauli-Hamiltonian the answer is less straightforward, as it depends on the
full parameter regime (involving also the magnetic field strength) but also
on the specific geometry of the fields. During certain geometric
configurations, the spin effects can be important in regimes of modest
density and modest temperature, which traditionally has been thought to be
completely classical \cite{Quant-classical}. A specific example of this kind
can be found in the MHD-regime. In Ref. \cite{Quant-classical} fluid
equations of the type (\ref{eq:mom-pauli})-(\ref{eq:spin}) was adopted to
MHD-regime, in order to study the physics of nonlinear spin-modified  Alfv\'en
waves. Within linear theory, the  Alfv\'en waves was almost unaffected by the
spin terms, provided the Zeeman energy associated with the unperturbed
magnetic field $B_{0}$ was much smaller than the thermal energy, i.e. $\mu
_{B}B_{0}\ll k_{B}T$. This condition holds for most plasmas except close to
pulsars and/or magnetars. However, nonlinearly the situation is different.
Ref. \cite{Quant-classical} used a two-fluid spin-model based on (\ref%
{eq:mom-pauli})-(\ref{eq:spin}), where spin-up and spin-down populations
were formally treated as different species, as a means to capture certain
kinetic effects within a more simple fluid theory. From this theory a
nonlinear Schr\"odinger equation%
\begin{equation}
i\partial _{t}B_{1}+\frac{v_{g}^{\prime }}{2}\partial _{\zeta }^{2}B_{1}+Q%
\frac{\left\vert B_{1}\right\vert ^{2}}{B_{0}^{2}}B_{1}=0  \label{eq:nls}
\end{equation}%
for  Alfv\'en waves propagating parallel to the magnetic field was derived,
where $B_{1}$ is the slowly varying magnetic field amplitude, $v_{g}^{\prime
}=dv_{g}/dk$ is the group dispersion, $\zeta =z-v_{g}t$ is the comoving
coordinate and $v_{g}$ is the group velocity. These quantities are
determined from the  Alfv\'en wave dispersion relation, which reads $\omega
^{2}=k^{2}c_{A}^{2}(1\pm kc_{A}/\omega _{ci})$, when weakly dispersive
effects due to the Hall current is included \cite{Cyclotron-ref}. Here $c_{A}
$ is the  Alfv\'en velocity and $\omega _{ci}$ is the ion-cyclotron frequency.
The upper (lower) sign corresponds to right (left) hand circular
polarization. The nonlinear coefficient is $Q=Q_{c}[1-({2}${$\mu _{B}B_{0}$}$%
/{m_{i}c_{A}^{2}})^{2}]$, where the classical coefficient is $%
Q_{c}=kc_{A}^{3}/4(c_{A}^{2}-c_{s}^{2})$ $\simeq -kc_{A}^{3}/4c_{s}^{2}$,
where $c_{s}$ is the ion-sound speed. Although linearly the modification of
the  Alfv\'en waves can be neglected (for modest temperature and densities),
the nonlinear coefficient $Q$ could be significantly affected by the spin
terms. Illustration of the parameters needed to make the different quantum
plasma effects significant, is shown in Fig. 1. In particular we note that
the two-fluid nonlinear spin effects are important for high plasma densities
and/or a weak (external) magnetic fields. For comparison, both the Fermi
pressure and the Bohm--de Broglie potential need a low-temperature or a very
high density to be significant.. A somewhat surprising result is that here
nonlinear spin effects tend to be more important for a lower magnetic field,
whereas the opposite is true for linear spin effects.

\begin{figure}
\includegraphics[width=0.8\columnwidth]{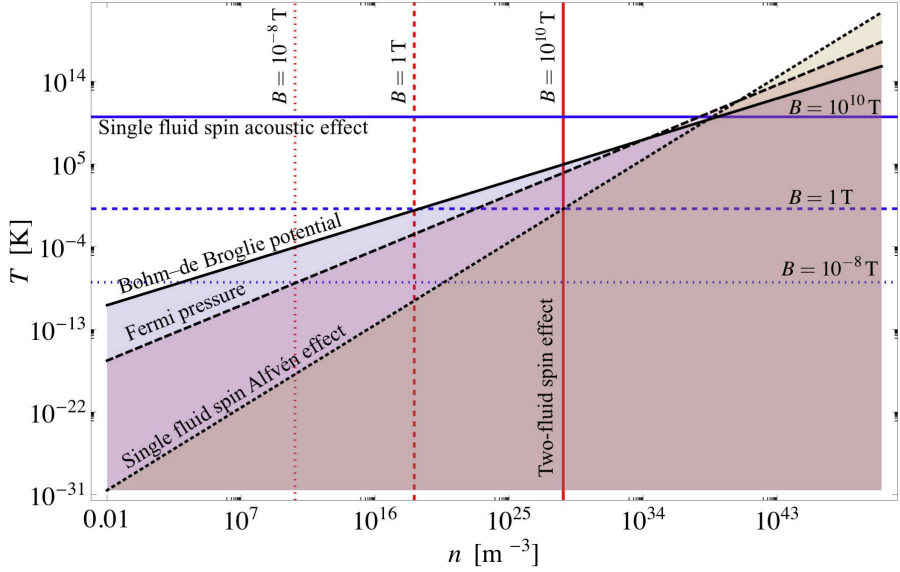}
\caption{\it A figure illustrating regions of importance in parameter space for
various quantum plasma effects. The lines are defined by different
dimensionless quantum parameters being equal to unity. The effects included
in the figure are: (i) Fermi pressure effects, described by the parameter $%
T_F/T_e \propto \hbar^{2}n_{0}^{2/3}/m k_{B}T_e$; (ii) Effects due to the
Bohm--de Broglie potential, described by the parameter $\hbar \protect\omega %
_{pe}/k_{B}T_e$; (iii) Spin single-fluid  Alfv\'enic effects, described by the
parameter $\hbar ^{2}\protect\omega _{pe}^{2}/mc^{2}k_{B}T_e$; (iv) Spin
single fluid acoustic effects, described by the parameter $\protect\mu%
_{B}B_{0}/k_{B}T_e$. Here lines for three different magnetic field strengths
are drawn. The quantum regime correspond to lower temperatures, i.e. it
exists below each of the three horizontal lines; (v) Spin two-fluid
nonlinear effects, described by the parameter $\protect\mu%
_{B}B_{0}/m_{i}c_{A}^{2}$. The quantum regime corresponds to higher
densities, i.e. it exists to the right of each of the three vertical lines.}
\end{figure}

As a second example of spin fluid effects we will consider the ponderomotive
force of an electromagnetic wave propagating along an external magnetic
field . Classically, the density fluctuations induced by the ponderomotive
force of an electromagnetic (EM) wave lead to an electrostatic wake field 
\cite{Wake-early}, as used in advanced particle accelerator schemes \cite%
{Adv. accelerator}. In other regimes, the back-reaction on the EM-wave due
to the density fluctuations leads to phenomena such as soliton formation,
self-focusing or wave collapse \cite{NL-effects,shukla86}. When spin effects
based on Eqs. (\ref{eq:mom-pauli})-(\ref{eq:spin}) are included, the
spin-contribution to the ponderomotive force, resulting from the combined
effect of the magnetic dipole force and spin-precession, leads to a
separation of spin-up and -down populations. The expression for the
ponderomotive force density for electromagnetic waves propagating along the
unperturbed magnetic field $\mathbf{B}_{0}=B_{0}\mathbf{z}$ can be divided
into its classical part 
\begin{equation}
F_{\mathrm{c}z}=-\frac{e^{2}}{2m^{2}\omega \left( \omega \pm \omega
_{c}\right) }\left[ \frac{\partial }{\partial z}\pm \frac{k\omega _{c}}{%
\omega \left( \omega \pm \omega _{c}\right) }\frac{\partial }{\partial t}%
\right] |E|^{2}.  \label{e10}
\end{equation}%
where $+(-)$ correspond to right (left) hand circular polarization, and the
spin part 
\begin{equation}
F_{\alpha z}=\mp \frac{2\mu ^{2}}{m\hbar }\frac{S_{0\alpha }}{\left( \omega
\pm \omega _{g}\right) }\left[ \frac{\partial }{\partial z}-\frac{k}{\left(
\omega \pm \omega _{g}\right) }\frac{\partial }{\partial t}\right] |B|^{2}.
\label{15}
\end{equation}%
\cite{brodinSpinPond}. Here $E$ and $B$ are the electric and magnetic field
amplitude, respectively, $\omega _{c}=eB_{0}/m$ is the cyclotron frequency, $%
\omega _{g}=2\mu B_{0}/\hbar =g\omega _{c}/2$. The index $\alpha =(u,d)$
refers to the up- and down populations, respectively. In particular the up-
and down spin vector components are $S_{0u}=1$, and $S_{0d}=-1$. Thus it is
clear that the spin-part of the ponderomotive force induces a separation of
the up- and down-populations. This effect survives also in the absence of an
external magnetic field, and it turns out that the magnitude of the relative
up-and down density perturbations can be larger than the classical density
perturbations in an unmagnetized plasma, provided%
\begin{equation}
1<\frac{\hbar \omega }{mc^{2}}\frac{\omega _{p}^{2}L^{2}}{c^{2}}
\label{condtion}
\end{equation}%
where $L$ is the pulse length of the high-frequency pulse. The first factor
of the right hand side of (\ref{condtion}) is smaller than unity for
frequencies below the Compton frequency, but the second factor can be large
for long pulse lengths, and hence large spin-polarization can be induced by
a sufficiently long EM-pulse in an unmagnetized plasma for optical
frequencies and higher. Once the plasma is spin-polarized, the spin terms in
the evolution equations can be important for the dynamics. For further
studies of results from spin-fluid theories, see e.g. Refs. \cite%
{Ferro,Spin-solitons}.

\subsection{Results from spin-kinetic theories}

The full kinetic theory (\ref{j8}) is accurate, but cumbersome for many
purposes. As seen from the full (\ref{j8}), the effects separates quite
naturally into particle dispersive effects (which are insignificant for
spatial scale lengths much longer than the characteristic de Broglie
wavelength), and effects due to the electron spin. The particle dispersive
effects has been studied in some detail in e.g. Refs. \cite%
{Manfredi2005,Shukla-Eliasson}. Focusing on the spin effects rather than
particle effects, we can therefore consider the long scale-length equation (%
\ref{j12}). An interesting effect of spin-kinetic theory is the appearance
of new wave-particle resonances. Linearized theory in a magnetized plasma
can be solved in much the same way as in a classical plasma (see e.g. Ref. 
\cite{brodin2008} for technical details). However, due to the fact that the
spin-precession frequency $\omega _{g}$ and the Larmor gyration occurs $%
\omega _{c}$ are slightly different (i.e. $\omega _{g}=(g/2)\omega
_{c}\approx 1.001\omega _{c}$) new resonances appear. In particular the
denominators of of the perturbed distribution function in kinetic theory are
replaced as 
\begin{equation}
\frac{1}{\left( \omega -k_{z}v_{z}-n\omega _{c}\right) }\rightarrow \frac{1}{%
\left( \omega -k_{z}v_{z}-n\omega _{c}-m\omega _{g}\right) }
\label{resonances}
\end{equation}%
when spin-kinetic effects are included, where the integer $n$ covers $\pm
\infty $ and $m=\pm 1$. Thus for a fixed parallel phase velocity, resonant
wave-particle interaction can occur at a much lower temperature when spin
effects are taken into account, since $\left\vert \omega _{c}-\omega
_{g}\right\vert \ll \left\vert \omega _{c}\right\vert $. This aspect has
been discussed in some detail in Ref. \cite{zamanian2010}. Furthermore, for
perpendicular propagation to the external magnetic field, new Bernstein-like
modes appear with frequencies close to the resonant value $\omega =\Delta
\omega _{c}=\left\vert \omega _{c}-\omega _{g}\right\vert $. Specifically
the dispersion relation for perpendicular propagation, reads%
\begin{eqnarray}
\omega ^{2} &=&\!\!\!k^{2}c^{2}+\omega _{p}^{2}\int \Big\{J_{0}^{2}\left( {%
k_{\bot }v_{\bot }}/{\omega _{c}}\right)   \notag \\
&&\!\!\!\!\!\!\!\!\!+\frac{k^{2}\hbar ^{2}\Delta \omega _{c}\sin ^{2}\theta
_{s}}{4m_{e}(\omega -\Delta \omega _{c})k_{B}T}\left[ J_{1}^{2}\left( {%
k_{\bot }v_{\bot }}/{\omega _{c}}\right) \right] \Big\}f_{0}\,d\Omega .
\label{Eq-DR}
\end{eqnarray}%
where $J_{0}$ and $J_{1}$ are zero and first order Bessel functions,
respectively. Here we have assumed the classical terms involving higher
order Bessel-functions are negligible, which is accurate for $\omega \ll
\left\vert \omega _{c}\right\vert $. The numerical solutions of (\ref{Eq-DR}%
) reveals that the wave frequency only deviates slightly from the resonance $%
\Delta \omega _{c}$ for most parameters \cite{brodin2008}. It should be
noted that a Madelung approach (c.f. Eqs (\ref{eq:mom-pauli})-(\ref{eq:spin}%
)) cannot capture the physics of the resonances at $\omega \approx \Delta
\omega _{c}$. However, the moment theory (\ref{cont-mom-exp})-(\ref{14})
correctly recovers the dispersion relation in the low temperature limit \cite%
{zamanian2010b}. Due to the higher order moments, however, it should be
noted that this theory is computationally somewhat more demanding than Eqs. (%
\ref{eq:mom-pauli})-(\ref{eq:spin}), at least if the higher order spin
effects are omitted in that theory. Other works on spin kinetic theory
includes Ref.\ \cite{Lundin2010}, where the general linear theory based on  (%
\ref{j12}) was studied in a magnetized plasma, Refs.\ \cite{cowley}, where a
simpler version of (\ref{j12}) (a semiclassical correspondence without the
magnetic dipole force term) was studied with regard to fusion applications,
and Ref.\ \cite{Asenjo2010} where a semiclassical version of (\ref{j12}) was
adopted to consider spin induced damping of electron plasma oscillations.

\section{Summary and Discussion}

A rapid development of spin-models for plasmas has taken place during the
last few years, using different types of methods. The most accurate are the
kinetic approach, based on the combined Wigner and q-transforms of the
density matrix. This theory captures not only the spin effects, but also the
particle dispersive effects. However, for scale lengths longer than the
characteristic de Broglie wavelengths only the spin effects remain, as
described by (\ref{j12}). While the kinetic approach is accurate, and
contains much interesting physics, as for example the resonances displayed
in Eq. (\ref{resonances}), there is a simultaneous need for models that are
simple enough to be applied also for more complicated problems, involving
inhomogenities and nonlinearities. Various fluid approaches have been
adopted for this purpose. Those based on the Madelung approach capture most
of the spin-physics in the physically intuitive effects of a magnetic dipole
force and spin precession, together with a spin magnetization current. Such
approaches has been used in several recent works, see e.g. \cite%
{Ferro,Spin-solitons,brodinSpinPond}. However, higher order quantum terms
also appear in this context, which either must be modelled or omitted, in
order to form a closed set for the fluid variables. Another approach to
obtain fluid theories is by computing moments of the kinetic theory. The
basic terms (i.e. spin-precession and the magnetic dipole force) are the
same as in the Madelung approach, but now a spin-velocity correlation tensor
appears in the evolution equation for the spin. This has a correspondence in
the tensor of (\ref{K-tensor}) in the Madelung approach, but here things are
complicated by the presence of other terms such as given by (\ref{Omega}).
Thus an advantage with the moment approach is that it is straightforward to
model the spin-velocity tensor by computing the corresponding moment. As
always when taking moments, the coupling to a higher moment appear, but
truncating the moment expansion in the next step (i.e. dropping higher order
moments in the evolution equation for the spin-velocity tensor) seem to
capture most basic spin physics, and leave out only thermal effects \cite%
{zamanian2010b}. However, even when one is interested in the low temperature
limit, certain effects of a nonzero temperature should be kept to address
the behavior in the more common plasma regimes. This has to do with the fact
that the Zeeman energy is typically much smaller than the thermal energy
(i.e. $\mu _{B}B_{0}\ll k_{B}T$), such that in thermodynamic equilibrium the
two spin states are almost equally populated. As a consequence, even if the
temperature is small in all other respects (i.e. all characteristic
velocities of a system is larger than the thermal (or Fermi) velocity),
there is a large spread in the spin distribution. As a means to capture some
of the physics associated with this large spread, two-fluid theories of
electrons has been developed where up- and down states with respect to the
(unperturbed) magnetic field are described as different species, as
described briefly in section 3.1. To some extent the fluid theories
including the spin-velocity tensor seem able to account for some of the
effects of the spread in the spin distribution, but it is too early to make
a definitive evaluation of the various models strengths and weaknesses. Thus
the final conclusion is that more research on the physics of electron spin
in plasmas is needed.


\begin{thebibliography}{99}

\bibitem{Shukla-Eliasson} P. K. Shukla and B. Eliasson,  Phys.\ Usp.\ 
\textbf{53}, 51 (2010).

\bibitem{holland} P.\ R.\ Holland, \textit{The Quantum Theory of Motion}
(Cambridge University Press, Cambridge, 1993).

\bibitem{dawson} J.\ Dawson, Phys.\ Fluids \textbf{4}, 869 (1961).

\bibitem{Pines} D.\ Pines, J. Nucl. Energy C: Plasma Phys. \textbf{2}, 5
(1961).

\bibitem{haas-etal1} F.\ Haas, G.\ Manfredi, and M.\ R.\ Feix, Phys.\ Rev.\
E \textbf{62}, 2763 (2000).

\bibitem{anderson-etal} D.\ Anderson, B.\ Hall, M.\ Lisak, and M.\ Marklund,
Phys.\ Rev.\ E \textbf{65}, 046417 (2002).

\bibitem{wigner} E.\ Wigner, Phys.\ Rev.\ \textbf{40}, 749 (1932).

\bibitem{moyal}  J.\ E.\ Moyal, Proc.\ Cambridge Philos.\ Soc.\  \textbf{45}%
, 99 (1949).

\bibitem{mendonca}  J. T. Mendonca, \textit{Theory of Photon Acceleration}
(IOP Publishing, 2001).

\bibitem{schleich}  W.\ P.\ Schleich, \textit{Quantum Optics in Phase Space}
(Wiley, 2001).

\bibitem{marklund} M.\ Marklund, Phys.\ Plasmas \textbf{12}, 082110 (2005).

\bibitem{Manfredi2005} G.\ Manfredi, Fields Inst.\ Comm.\  \textbf{46}, 263
(2005).

\bibitem{marklund-brodin} M.\ Marklund and G.\ Brodin, Phys.\ Rev.\ Lett.\  
\textbf{98}, 025001 (2007).

\bibitem{brodin-marklund} G.\ Brodin and M.\ Marklund, New J.\ Phys.\  
\textbf{9}, 277 (2007).

\bibitem{degroot-suttorp} S.\ R.\ de Groot and L.\ G.\ Suttorp,  \textit{%
Foundations of Electrodynamics} (North-Holland, 1972).

\bibitem{alfven} H.\ Alfv\'en, Nature \textbf{150}, 405 (1942).

\bibitem{halperin-hohenberg} B.\ I.\ Halperin and P.\ C.\ Hohenberg, Phys.\
Rev.\  \textbf{188}, 898 (1969).

\bibitem{balatsky} A.\ V.\ Balatsky, Phys.\ Rev.\ B \textbf{42}, 8103 (1990).

\bibitem{rathe-etal} U. W. Rathe, C. H. Keitel, M. Protopapas, and P. L.
Knight,  J. Phys. B: At. Mol. Opt. Phys. \textbf{30}, L531 (1997).

\bibitem{hu-keitel} S. X. Hu and C. H. Keitel, Phys. Rev. Lett. \textbf{83},
4709 (1999).

\bibitem{arvieu-etal} R.\ Arvieu, P.\ Rozmej, and M.\ Turek, Phys.\ Rev.\ A  
\textbf{62}, 022514 (2000).

\bibitem{aldana-roso} J.\ R.\ V\'azquez de Aldana and L.\ Roso, J.\ Phys.\
B:  At.\ Mol.\ Opt.\ Phys.\ \textbf{33}, 3701 (2000).

\bibitem{walser-keitel} M. W. Walser and C. H. Keitel, J. Phys. B: At. Mol. 
Opt. Phys. \textbf{33}, L221 (2000).

\bibitem{qian-vignale} Z. Qian and G. Vignale, Phys. Rev. Lett. \textbf{88},
056404 (2002).

\bibitem{walser-etal} M.\ W.\ Walser, D.\ J.\ Urbach, K.\ Z.\ Hatsagortsyan,
S.\  X.\ Hu, and C.\ H.\ Keitel, Phys.\ Rev.\ A \textbf{65}, 043410 (2002).

\bibitem{roman-etal} J.\ S.\ Roman, L.\ Roso, and L.\ Plaja, J.\ Phys.\ B:
At.\  Mol.\ Opt.\ Phys.\ \textbf{37}, 435 (2004).


\bibitem{liboff} R.\ L.\ Liboff, Europhys.\ Lett. \textbf{68}, 577 (2004).

\bibitem{fuchs-etal} J.\ N.\ Fuchs, D.\ M.\ Gangardt, T.\ Keilman, and G.\
V.\  Shlyapnikov, Phys.\ Rev.\ Lett.\ \textbf{95}, 150402 (2005).

\bibitem{kirsebom-etal} K.\ Kirsebom \textit{et al.}, Phys.\ Rev.\ Lett.\  
\textbf{87}, 054801 (2001).

\bibitem{weyl} H.\ Weyl, \textit{The Theory of Groups and Quantum Mechanics}
(Dover, New York, 1931).

\bibitem{glauber2} R.\ J.\ Glauber, Phys.\ Rev.\ \textbf{131}, 2766 (1963).

\bibitem{leonhardt} U.\ Leonhardt, \textit{Measuring the Quantum State of
Light},  (Cambridge University Press, Cambridge 1997).

\bibitem{lee2} H.\ W.\ Lee, and M.\ O\. Scully, Found.\ Phys.\ \textbf{13},
61 (1983).

\bibitem{carruthers} P.\ Carruthers, and F.\ Zachariasen, Rev.\ Mod.\ Phys.\
\textbf{55}, 245 (1983).

\bibitem{takahashi} K.\ Takahashi, Prog.\ Theor.\ Phys.\ Supp.\ \textbf{98},
109 (1989).

\bibitem{haug} H.\ Haug, and A-P Jauho  \textit{Quantum Kinetics in
Transport and Optics of Semiconductors} (Springer Series in  Solid-State
Sciences vol 123) (Berlin: Springer 2007).

\bibitem{rammer} J.\ Rammer, \textit{Quantum Transport Theory} (Cambridge: 
Cambridge University Press 2007).

\bibitem{glauber} R.\ J.\ Glauber, Phys.\ Rev.\ \textbf{130}, 2529 (1963).

\bibitem{sudarshan} E.\ C.\ G.\ Sudarshan, Phys.\ Rev.\ Lett.\ \textbf{10},
277 (1963).

\bibitem{glauber65} R.\ J.\ Glauber, \textit{Quantum Optics and Electronics}
edited by C.\ Dewitt, A.\ Blandin and C.\ Cohen-Tannoudji (Gordon and
Breach, New York,  1965).

\bibitem{husimi} K.\ Husimi, Prog.\ Phys.\ Math.\ Soc.\ Japan \textbf{22},
264 (1940).

\bibitem{lee} H.\ W.\ Lee, Phys.\ Rep.\ \textbf{259}, 147 (1995).

\bibitem{stratonovich} R.\ L.\ Stratonovich, Sov.\ Phys.\ D \textbf{1}, 414
(1956).

\bibitem{serimaa} O.\ Serimaa, J.\ Javanainen, and S.\ Varro,  Phys.\ Rev.\
A \textbf{33}, 2913 (1986).

\bibitem{cohen} L.\ Cohen, and M.\ Scully, Found.\ Phys.\ \textbf{16}, 295
(1986). 

\bibitem{chandler} C.\ Chandler, L.\ Cohen, C.\ Lee, and M.\ Scully, 
Found.\ Phys.\ \textbf{22}, 267 (1992).

\bibitem{scully} M.\ Scully, and K.\ Wodkiewicz,  Found.\ Phys.\ \textbf{24}%
, 85 (1994).

\bibitem{cunha} M.\ Cunha, V.\ Man'ko, and M.\ Scully, Found.\ Phys.\ Lett.\
\textbf{14}, 103 (2001).

\bibitem{zamanian2010} J.\ Zamanian, M.\ Marklund, and G.\ Brodin,  New J.\
Phys.\ \textbf{12}, 043019 (2010).

\bibitem{schwinger} J.\ Schwinger, Phys.\ Rev.\ \textbf{73}, 416 (1948).

\bibitem{brodin2008} G.\ Brodin, M.\ Marklund, J.\ Zamanian, A.\ Ericsson, 
and L.\ P.\ Mana, Phys.\ Rev.\ Lett.\ \textbf{101}, 245002 (2008).

\bibitem{cowley} S.\ C. Cowley, R. M. Kulsrud, and E. Valeo, Phys. Fluids 
\textbf{29}, 430 (1986).

\bibitem{nicholson} D.\ Nicholson, \textit{Introduction to Plasma Theory}, 
(John Wiley \& Sons Inc 1983).

\bibitem{haas0} F. Haas, Phys. Plasmas \textbf{12}, 062117 (2005).

\bibitem{haas1} F. Haas, M. Marklund, G. Brodin, and J. Zamanian, Phys. 
Lett. A \textbf{374}, 481 (2010).

\bibitem{haas2} F. Haas, J. Zamanian, M. Marklund, and G. Brodin,  New. J.
Phys. \textbf{12}, 073027 (2010).

\bibitem{zamanian2010b} J. Zamanian, M. Stefan, M. Marklund, and G. Brodin, 
Phys. Plasma, in press (2010).

\bibitem{zhang} W. Zhang, and R. Balescu, J. Plasma Phys. \textbf{40}, 199
(1988).

\bibitem{balescu} R. Balescu, and W. Zhang, J. Plasma Phys. 40, 215 (1988).

\bibitem{Quant-classical} G. Brodin, M. Marklund and G. Manfredi, Phys. Rev.
Lett. \textbf{100}, 175001 (2008).

\bibitem{Cyclotron-ref} G. Brodin and L. Stenflo, Contrib. Plasma Phys. 
\textbf{30}, 413 (1990).

\bibitem{Wake-early} L. M. Gorbunov and V. I. Kirsanov, Zh. Eksp. Teor. Fiz.
\textbf{93}, 509 (1987) [Sov. J. Plasma Phys. \textbf{93}, 290 (1987)].

\bibitem{Adv. accelerator} R. Bingham, Nature \textbf{445}, 721 (2007).

\bibitem{NL-effects} L. Berge, Phys. Reports \textbf{303} 259 (1998)

\bibitem{shukla86} P. K. Shukla, N. N. Rao, M. Y. Yu and N. L. Tsintsadze, 
Phys. Reports, \textbf{138}, 1 (1986).

\bibitem{Ferro} G. Brodin and M. Marklund,  Phys. Rev. E \textbf{76}, 055403
(2007).

\bibitem{Spin-solitons} G. Brodin and M. Marklund,  Phys. Plasmas \textbf{14}%
, 112107 (2007).

\bibitem{brodinSpinPond} G. Brodin, A. P. Misra, M. Marklund,  Phys. Rev.
Lett. \textbf{105}, 105004 (2010)

\bibitem{Asenjo2010} P. S. Moya and F. A Asenjo, arXiv 1005.2573
[physics-plasm-ph]

\bibitem{Lundin2010} J. Lundin and G. Brodin, arXiv:1006.1310v1
[physics.plasm-ph]
\end{thebibliography}
\end{document}